\documentclass[11pt]{amsart}
\usepackage{geometry}                
\usepackage{graphicx}
\usepackage{amssymb}
\usepackage{epstopdf}
\title{The Physics Behind Symmetrization}
                                     
\begin{document}
\maketitle

\centerline{R. E. Kastner, University of Maryland, College Park. rkastner@umd.edu}
\centerline{20 June 2026}\smallskip

ABSTRACT. It is often asserted that quantum states for same-type particles must be symmetrized due to ``label redundancy,'' i.e. the assumption
that the permutations of labels in direct-product states do not reflect any real physical distinction and thus their permutations constitute
an ``exchange degeneracy''. This assumption is directly challenged by the case
of scattering of same-type particles such as electrons, which involves two physically distinct scattering channels effectively corresponding to permutation
of the labels.  I discuss this counterexample with critical attention to an extant portrayal in the literature that omits pertinent physical content. I further note ways in which the assumption that symmetrization must be universally imposed is not supported by actual calculations of particle interactions, nor by seemingly viable particle states based on preparations and outcomes. 
\smallskip


\section{Introduction and Background}

This paper addresses the interpretation of symmetrized states in quantum theory. Symmetrized states have long been known to be empirically necessary for correct predictions involving certain multi-particle systems of the same type. For example, the two electrons in a Helium atom occupy eigenstates which are sums of direct products of individual particle Hilbert space states in which arbitrary labels of the particles (e.g., 1 and 2) have been permuted. These are termed ``singlet states'' or ``triplet states'' depending on the collective angular momentum value. For fermions, the total (spin + space) states must be ``antisymmetrized'' (have opposite phases) while for bosons, the total states must be symmetrized (have the same phase). The spin singlet state for electrons is:

$$|\Psi_{singlet}\rangle = \frac{1}{\sqrt 2} (|\uparrow_1\rangle \otimes |\downarrow_2\rangle - |\uparrow_2\rangle \otimes |\downarrow_1\rangle) \eqno(1)$$

 We should first note that a key issue concerns exactly what constitutes a ``multi-particle system'', since it is acknowledged that in the real world, particles
 of the same type often do not exhibit correlations signifying entanglement (or, more generally, a requirement for explicit symmetrization).\footnote{An answer to the question of ``what constitutes a multi-particle system'' for symmetrization purposes is offered in Kastner (2023).} In other words, it seems empirically valid for same-type particles to be not formally symmetrized under certain circumstances. This remains an ill-defined situation in conventional quantum theory, primarily due to the measurement problem. The latter gives rise to ambiguities about whether or how a given quantum system may be accurately described simply by an eigenstate reflecting an outcome for the measured observable--as, for example, a single electron being found (or prepared) in the state $|p, z+\rangle$. The latter is clearly not a symmetrized state, yet such state descriptions are used endemically without empirical inconsistency. For example, one routinely conducts Stern-Gerlach experiments with individual electrons in some prepared state like the above, where that state is not symmetrized over all other electrons in the universe. 
 
 There is additional ambiguity concerning whether formally symmetrized states are entangled. Such a determination rests on the definitions of these terms. Some authors, such as Schlichtholz and Markiewicz (2023), refer to symmetrized states as ``entangled'' (in a formal sense) but deny that such formal entanglement results in ``usable'' entanglement on the basis that one cannot construct appropriate local observables. They say:
 \begin{quote}
  Different types of entanglement were considered in the Fock space...One of them is the concept of entanglement of indistinguishable particles emerging
from the symmetrization procedure in Fock space construction. However, let us comment that this type
of entanglement is not directly accessible by nature and rather seems to be a mathematical artifact of the
construction of the Fock space. This is because one cannot build local observables without some effective distinguishability. 
\end{quote}
 
 For our current purposes, we need take no position on this issue of discerning symmetrization from functional entanglement, which is peripheral to our main focus:  specifically, whether there is real redundancy in particle labels and how the answer to that question affects the status of the so-called ``symmetrization postulate''. In short, we find that symmetrization is not something that needs to be postulated as a universal rule solely in order to avoid redundancies and associated inconsistencies, but instead is a procedure demanded by specific physics and that describes specific physics. In fact, no genuine redundancies arise when neglected physics is appropriately taken into account. Moreover, we argue that unnecessary redundancies, or even inconsistencies, may be introduced through treating symmetrization as a  universal postulate. 
 
Before continuing, we note that this paper addresses primarily fermion symmetrization. We discuss boson symmetrization in Section 5.   

  The prevailing way of attempting to justify symmetrization in terms of a universal postulate is exemplified by the following passage in Bigaj (2015) which invokes ``exchange degeneracy'' or the notion of label redundancy (emphasis added for critical purposes): 

\begin{quote}

\small{``The textbook way to introduce this postulate is through the concept of exchange degeneracy [Cohen-Tannoudji, C., Diu, B., $\&$ Laloe, F. (1977)].  Considering the joint state of two particles of the same type such that one of them occupies state $|u\rangle$ whereas the other one is in a different state $| v\rangle$, {\bf we should observe that the two permuted states $    | {\mathbf u}\rangle |{\mathbf v} \rangle$ and $|{\mathbf v} \rangle |{\mathbf u} \rangle $ are empirically indistinguishable}. According to the essentialist approach this indistinguishability comes from the fact that both bi-partite states represent one and the same physical state of affairs. On the other hand, the haecceitist approach admits that there is a difference between the permuted and non-permuted states, but this difference cannot give rise to any observational effects, as haecceities are not empirically accessible. In order to avoid the degeneracy problem, we adopt the symmetrization postulate, which narrows down the admissible states to the symmetric (occupied by bosons) and antisymmetric ones (applicable to fermions).'' }

\end{quote} 

 We will question the bolded passage in Section 3. For now, we observe that Bigaj advocates what he terms the ``essentialist'' approach which asserts a true label redundancy, while he mentions also a competing ``haecceitist'' approach according to which the distinction referred to by the permutation is viewed as not related to specific physics but only to a metaphysical principle of fundamental `thisness'. 
 In keeping with his view that particle labels are truly redundant, Bigaj has advocated an approach that rejects so-called ``factorism'', Caulton's term (e.g., Caulton 2014) for the idea that the individual direct product states appearing in symmetrized states are physically meaningful.
 
  In direct contrast to this position, the current work specifically defends the meaningfulness of the direct-product states. (We do not however adopt the term ``factorism'' to describe this position in view of the term's pejorative associations. For example,  Caulton claims that ``factorist'' states cannot satisfy the classical limit (Caulton 2014, 14). However, that conclusion is based on failure to take into account a well-defined measurement process and is directly contested in Kastner, 2023 on that basis. Dieks and Lubberdink (2022) also reject physical meaningfulness of the direct product states).
  
 We address the above interpretive issues further in Sections 3 and 4. We now turn to a specific counterexample to the assumption of label degeneracy exemplified in the Cohen-Tannoudji {\it et al} extract quoted above. The example calls into question the basic ``ground rules'' on both sides of the conventional debate (essentialist vs. haecceitist), both of which deny any specific physical significance associated with the label permutation.

\section{The scattering example}

The ``anti-factorism'' view is a deflationary approach in that it denies that there is any physical meaning in individual system spaces and their direct products. This approach is exemplified in the portrayal in Bigaj (2022, 224-7) of electron scattering as involving only label-swapping, when in fact there is physical content that, when represented in a Feynman diagram, would readily distinguish the two types of processes in play. We now examine these details.

The author considers electron-electron (M\o ller) scattering but does not provide the associated tree-level Feynman diagrams, so let us provide those first for reference in Figure 1. They are called the ``t-channel'' and the ``u-channel''. The former involves simply connected paths from initial points to final points, while the latter involves path crossing, and the u-channel is also called the ``crossing channel'' for that reason.

\begin{figure}[!h]
\centering\includegraphics[width=1.5in]{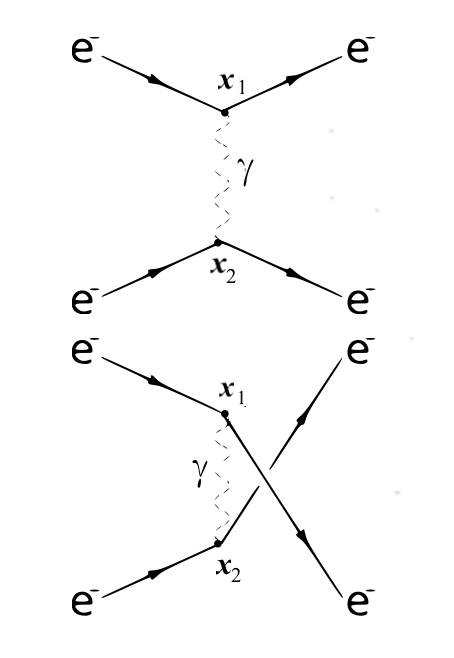}

\label{fig_sim}

  \caption{Feynman diagrams for M\o ller scattering t- and u-channels, top and bottom respectively}

\label{fig_sim}
\end{figure}

We should first note that, contrary to the presentation in Bigaj (2022, 226) that ascribes an entangled (singlet) momentum state to the incoming electrons, the correct Feynman-diagram analysis of scattering interactions treats incoming particles as independent in an ``appropriate limit'' which is usually codified as $t \rightarrow  -\infty$. In the appropriate limit, there is a fact of the matter, empirically grounded, that ``electron prepared in momentum state $p_1$ (or $p_2$) has spin $i$ (or spin $j$) .''
Thus, the incoming electrons are factorizable, whether or not symmetrization is imposed as a formal procedure. In that sense, it is arguably redundant; we deal with this issue further below and in Section 4. The presentation in Bigaj (2022) further departs from the actual physics of the scattering situation by omitting the interaction Hamiltonian that is always present for electrons, using instead a trivial Hamiltonian. Thus, his obtained outgoing (final state) interference arises only through the assumption of an incoming entangled state, which is not the case in general. Instead, interference arises from the interaction Hamiltonian; i.e., it is what correlates the independent incoming electrons. (This means that if one also assumes 
 well-defined single-particle outgoing states, an implicit post-interaction measurement is involved. This subtlety is typically glossed in an ambiguity concerning whether the outgoing momenta and spin $p_i',s_j$ are variables or constants. If taken as constants, a measurement resulting in a well-defined single electron spinor states is implied. If taken as variables, they remain correlated in an antisymmetric state with total momentum and spin subject to conservation laws.) The point of this observation is to underscore the fact that physically relevant symmetrization--as in the symmetrized and functionally entangled total space + spin state assumed in Bigaj's treatment--is not present universally but arises due to specific physical processes, including interaction Hamiltonians. While the exchange interaction is often not made explicit, that is what gives rise to two interfering scattering channels. It is described by the exchange integral (cf. Hutem and Boonchui, 2012): 

 $$K_{1,2} = \int_{\bf{r_1}} \int_{\bf {r_2}}  \psi^*_a({\bf r_1})  \psi^*_b {(\bf r_2})  \frac{e^2}{| {\bf r_2} - \bf{r_1}|} 
    \psi_a ({\bf{r_2}})  \psi_b ({\bf{r_1}})  d {\bf {r_1}} d \bf{r_2} \eqno(2)  $$ \smallskip
    
\noindent  for $\{a,b\}$ designating two different possible spatial states for the particles and $\{1,2\}$ two arbitrary labels for the particles.  Thus it is a specific interaction (or process) that imposes no fact of the matter about which electron is in which state ($a$ or $b$), not a formal postulate.  Since this point is rather crucial, let us look at the explicit expressions for the scattering amplitudes, which reflect the independence of the incoming states and the actual source of the resulting symmetrization.  (In the following we drop common factors including masses, energies, and a conserving delta function; $u_r({\bf p}_i)$ are spinors where $r_i$=\{up, down\} along some specified axis, $\bar{u}_{r'}({\bf p}'_i)$ are conjugate (outgoing) spinors, and $D_{F_{\mu \nu}}$ is the Feynman photon propagator with factor $g_{\mu \nu}$ compactly indicated through the greek indices). The amplitude for the t-channel, where ``$p_1$ goes to $p'_1$ and $p_2$ goes to $p'_2$'' is:

$$S^{(t-channel)} = -D_F(p'_1-p_1)_{\mu \nu}\  \bar{u}_{r_1'}({\bf p}'_{1})\gamma^\mu u_{r_1}({\bf p}_{1}) \bar{u}_{r_2'}({\bf p}'_{2})\gamma^\nu u_{r_2}({\bf p}_{2}) \eqno(3)$$

And the amplitude for the u-channel, where ``$p_1$ goes to $p'_2$" and $p_2$ goes to $p'_1$'' is:

$$S^{(u-channel)} = D_F(p'_2-p_1)_{\mu \nu}\  \bar{u}_{r_2'}({\bf p}'_{2})\gamma^\mu u_{r_1}({\bf p}_{1}) \bar{u}_{r_2'}({\bf p}'_{1})\gamma^\nu u_{r_2}({\bf p}_{2}) \eqno(4)$$
\smallskip
 
It is these two distinct channels that correspond to the effective antisymmetrization of the electrons, not a formal exchange of individual Hilbert space labels in the absence of any relevant physical process.  That is, permuting the relation of incoming momenta to outgoing momenta transforms the t-channel to the u-channel. The latter corresponds to a kind of ``label permutation,'' but only insofar as the individual Hilbert space labels act as bookkeeping devices for the two channels.

Of course, conventionally one treats the incoming electrons as a Fock state, but the amplitude calculation via the S-matrix involves two distinct internal points $x_1$ and $x_2$ (labeling the two vertices in the Feynman diagrams) where the field operator $\psi(x)$ acts to transform one electron state into another at that point. Crucially, the field operator acts at that particular point  $x_i$  and only the incoming particular field state at that vertex survives to be transformed into an outgoing states; all other states are destroyed. To see this, first recall the basic form of the Dirac electron fields:

\begin{figure}[!h]
\centering\includegraphics[width=3in]{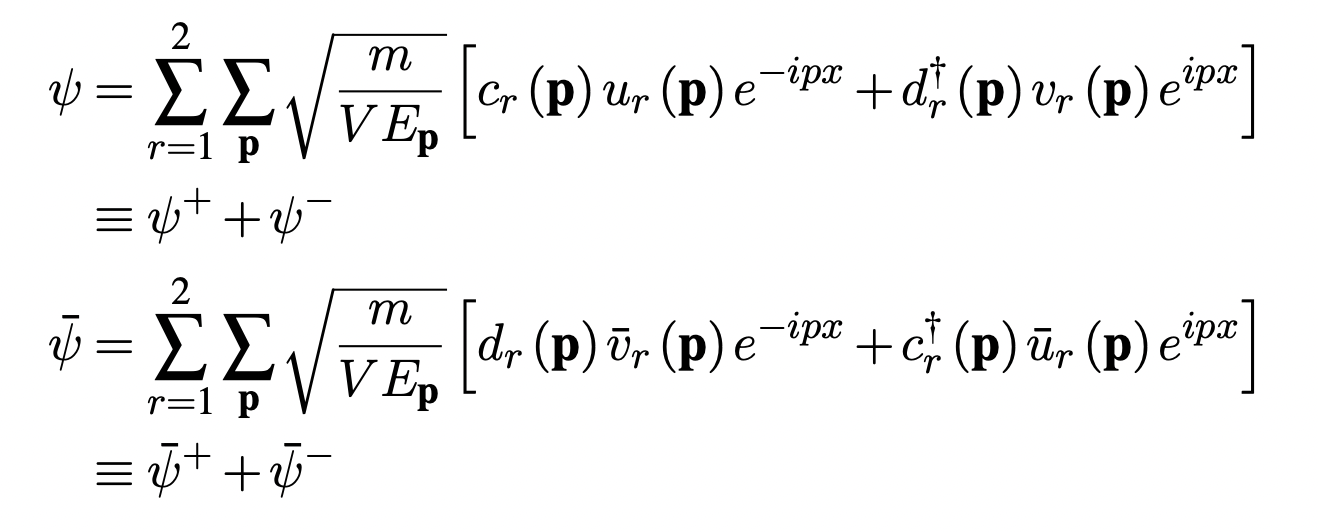}



\end{figure}

These field operators appear in the relevant term of the S-matrix for each channel/amplitude. In particular, the S-matrix 
(second order) term for the T-channel is: 

$$S^2_{M1} = i e^2 \int d^4 x_1 d^4 x_2 D_{F,\mu\nu} (x_2 - x_1) \bar{\psi}_{1'}^{-} (x_1) \gamma^\mu \bar{\psi}_{2'}^{-} (x_2)
\psi_{1}^{+} (x_1) \gamma^\nu \psi_{2}^{+} (x_2)\eqno(T)$$

\noindent where $D_F$ is the Feynman propagator and the spinor operators are ``normal-ordered'' such that the annihilation operators (e.g., $\psi_{1}^{+} (x_1)$ are on the right and the creation operators (e.g.,$ \bar{\psi}_{1'}^{-} (x_1)$)  are on the left. This entire expression will act on the incoming state, with the field as defined above as sums of the field operators for the possible states. The annihilation operators act on the incoming state $| 1_{p_1, r_1}, 1_{p_2, r_2}\rangle$ at vertices $x_1$ and $x_2$. Thus,
 the only terms in above sums that survive are those that match the incoming or outgoing momentum and spin at vertex $x_1$ or $x_2$.  Since one coordinate can only be associated with one momentum value at a time, that coordinate essentially replaces an arbitrary individual Hilbert space label for any particular diagram (channel). Thus the two channels override any initial formal symmetrization.  In other words, the relevant S matrix term for each scattering channel converts any initial formal symmetrization to a single particular state associated with each vertex, as we can see from the results (3) and (4) for the t-channel and u-channel, respectively. This means that the initial symmetrization is doing no actual computational work; instead it is the two scattering amplitudes that express the relevant symmetrization, and that arises via the action of the field operators.
 
  The point is that there has to be some physical context demanding that the field operators act on a specific state in order for symmetrization to reflect the commutation relations of the field operators. The fact of that commutation relation in itself does not rise to the level of some intrinsic symmetry of the particles themselves, independently of the action of the field operators. In particular, the relative minus sign for the u-channel comes not from any pre-existing abstract symmetry property of the incoming electrons, but from the need to exchange the spinor operators  $\bar{\psi}_{1'}^{-} (x_i)$ and $ \bar{\psi}_{2'}^{-} (x_j)$ to get from the t-channel process, in which the incoming momenta are transformed to their primed versions, to the u-channel process, in which there is ``crossing'' such that $p_1$ is transformed into $p_2'$ and vice versa. 


  Interestingly, there is a duplicate diagram for each channel where the labels $x_1$ and $x_2$ are interchanged, since there is no fact of the matter about `which electron is at which vertex'; but crucially, that interchanging of the labels of the vertices is not ``fermion symmetrization'' since the resulting amplitudes have the same sign and these sub-amplitudes add, doubling the amplitude of each channel. Thus, it is the need to sum the {\it distinct amplitudes for each channel} that creates the effective antisymmetrization and resulting interference, not a prior formal symmetrization of the state of the electrons (which, even if assumed, is made irrelevant by the action of the field operators on specific states at specific designated coordinates $x_i$, as we see above). Instead it is the permutation of the field operators due to the differing momentum states emerging at the vertices, and the need to sum those individual amplitudes, that corresponds to the fermionic antisymmetrization. Thus the antisymmetry results from permutation of the field operators acting on specific outgoing momentum/spin states, not global permutation of particle labels throughout the interaction.

The ``indistinguishability'' manifesting in (2) arises from the fact that electrons are all excitations of the same quantum field, whose operators are destroying the incoming electron states and creating the outgoing ones at the vertices. The field does not account for ``which electron is in state'', and that is why there are two possible ways for the incoming momenta to transition to the outgoing momenta, corresponding to the amplitudes (3) and (4). But lest this be taken as amounting to some formal reason for symmetrization, it is important to note that the interference effect of M\o ller scattering reflecting symmetrization arises due to a specific interaction--and not without its operation.
 
 Indeed, Bigaj (apparently inadvertently) illustrates this exact point when he includes the spin degrees of freedom in his incoming formally symmetrized electron state and shows that the interference disappears under the trivial Hamiltonian (Bigaj 2022, 228-9). Thus, the formal symmetrization of two electrons assumed to be in specific orthogonal states and subject only to a trivial Hamiltonian demonstrably fails to account for the observed behaviors of electrons. Instead, it is the interaction (2) manifesting in the two amplitudes (3) and (4) that effects the physical correlation between the electrons. One finds, therefore, that it is it neither necessary nor sufficient to impose symmetrization as a universal formal rule, nor to invoke a redundancy with respect to permutation. As we saw with respect to the quantitative details of the channels, in fact there is no such redundancy.
 
 The key point is that the only justification for symmetrization is in terms of the commutation relations of the field operators, not any formal label exchange. This is borne out by the fact that there is no sign change for the label permutation involved in exchanging the vertex labels (thus doubling each diagram's contribution). The only exchange leading to a sign change comes from exchanging which momentum state is being created at each vertex.  The latter sort of process also applies to the pair correlation function $g^{(2)}$ for the Dirac field (cf. Eisert, 2024), but this is only short-range and is legitimately neglected for M\o ller scattering. Indeed if we look at the leading (zero interaction) term of the S-matrix, this is routinely omitted since the formally antisymmetrized electrons pass by one another with no scattering at all; see, e.g. Beisert (2026), p.9.5.
  It should also be emphasized that any pair correlation contribution arises from the physically relevant permutation of field operators in a particular context. Clearly that has no significant contribution to M\o ller scattering, and thus cannot be invoked as a justification for symmetrization in the context of M\o ller scattering. It is of course  true that $a^\dag b^\dag |0> = - b^\dag a^\dag |0>$ , which is often used as the justification for assigning antisymmetric states universally to fermions (and mutatis mutandis for bosons, with a plus sign corresponding to symmetric states).  However, the commutation relations only enter into symmetrization in the context of specific physical processes. Thus, while Fock states are a useful basis for quantum field theory, they don't necessarily reflect the physically relevant details of interactions, such as scattering processes or higher-order emissions and absorptions of photons. 

 Let us further note the physical distinction between the two M\o ller channels in terms of their basic topological structures, which are obscured in the  ``center of mass'' depiction in Bigaj (2022) of the two tree-level processes. That depiction obscures the fact that the u-channel involves crossing, as can be seen in a reproduction of his Figure 8.1 (our Figure 2). The author's non-intersecting depiction is first used in reference to distinguishable particles, but he then discusses particles of the same type (``indistinguishable particles'') with reference to the same figure. In Figure 3, we focus on the u-channel as sketched in his left-hand diagram. In the author's depiction, any lab-frame longitudinal propagation is suppressed, as is temporal dependence. This obscures significant physical content, which we now consider with reference to Figure 4.

\begin{figure}[h]	
   \centering
    \includegraphics[width=0.4\textwidth]{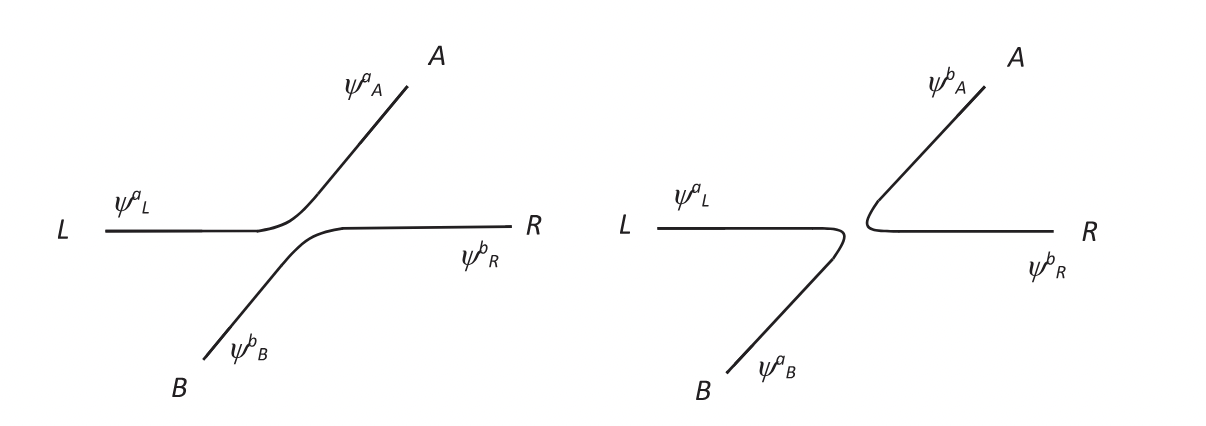}
    \caption{Bigaj's Figure 8.1}
    \label{fig:your-label}
\end{figure}


\begin{figure}[h]
   \centering
    \includegraphics[width=0.25\textwidth]{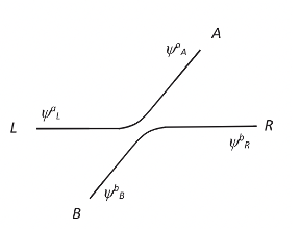}
    \caption{Bigaj's Figure 8.1, first image.}
    \label{fig:your-label}
\end{figure}

\begin{figure}[h]
   \centering
     \includegraphics[width=0.5\textwidth]{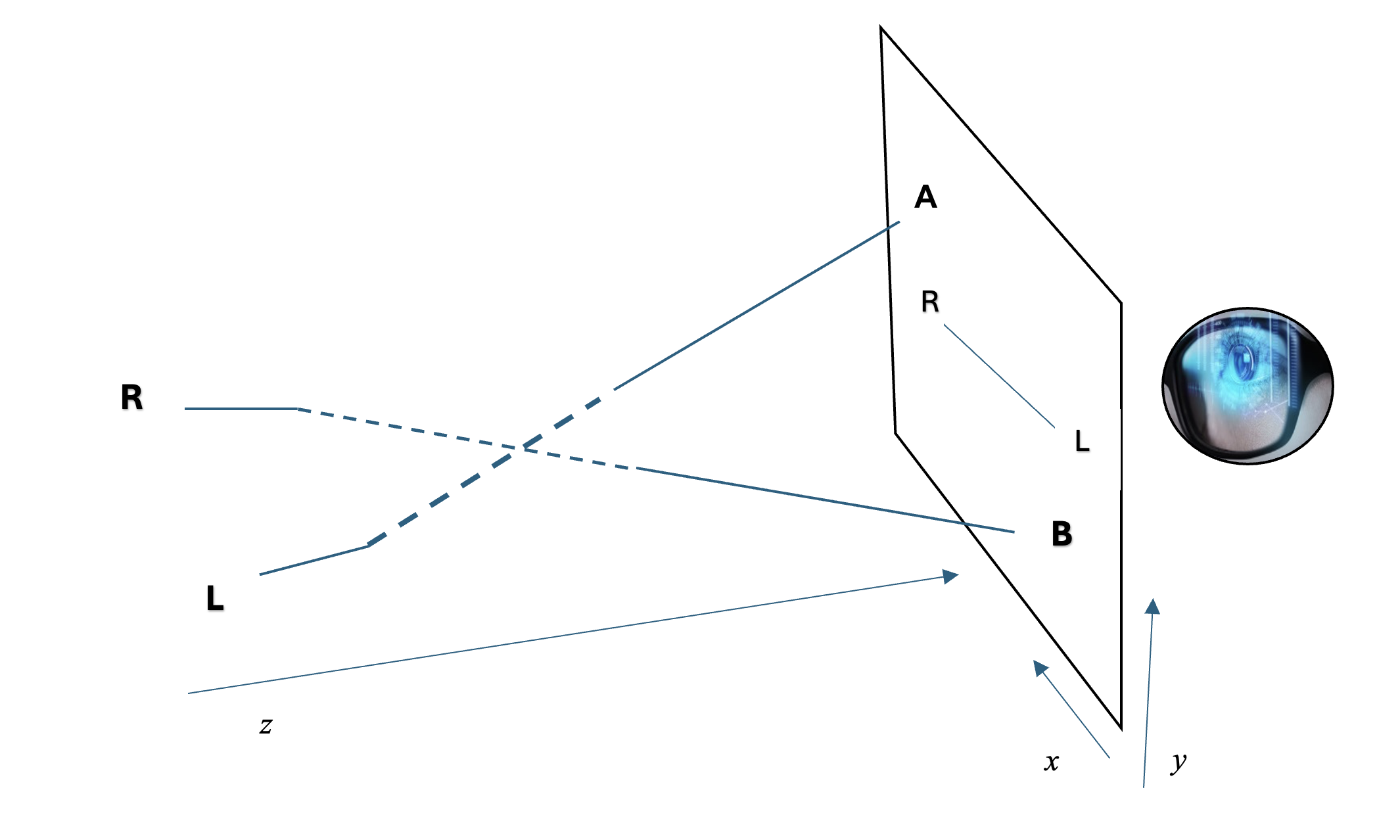}
    \caption{Bigaj's depiction from a different viewpoint.}
    \label{fig:your-label}
\end{figure}

The depicted R  and L electrons wavefunction propagation can be readily seen as a function of time if we parameterize the $z$ (longitudinal) component by $z(t) = t$. (The propagation may not necessarily have a z component, but this helps to visualize the t-dependence, as in a Feynman-type diagram). Then Bigaj's depiction corresponds to the viewpoint of the indicated observer shown here in Figure 4; the observer does not `see' the z-propagation.

 In the author's u-channel diagram, Figure 3, the electrons are shown as propagating towards one another in the x,z plane and then veering off in the plus and minus y directions respectively, never intersecting. (From our viewpoint looking down at Figure 4, the L trajectory is in front of the R trajectory).  But in fact, this is a ``crossing'' or u-channel process that does not preclude the trajectories from intersecting; that is why it is called a ``crossing'' channel. 
In particular, if the beginning and ending points are all in the same plane, then the paths must cross. Specifically, let A and B be at the same y-coordinates as the initial states R and L (as illustrated by their indicated projections on the detection plane in Figure 4). Then the trajectories would have to intersect when illustrated in the same manner as Bigaj's figure 8.1; i.e., there could be no veering off to different y-values. This is essentially what is depicted in the Feynman diagram for the u-channel. Furthermore, if the propagation had no z-component (i.e., contained in the xy plane), for endpoints A and B at the same y-coordinate, in the u-channel the particle paths would intersect identically throughout.\footnote{In using the terminology ``path'' or ``trajectory'' we keep in mind that these are {\it possible} paths, as in Feynman sum-over-paths. These are not determinate classical trajectories. Nevertheless, they are individually physically well-defined in terms of their respective momenta. We might also observe that while one could deny that either trajectory is really continuous and thus evade the intersecting aspect of the topological distinction, this does not nullify the observation that each permutation describes a physically distinct type of connection between incoming and outgoing momenta.} 

 In summary, the t-channel and the u-channel have distinct names because they are topologically distinct physical processes. The
two cases depicted by the author as representing nothing more than a swapping of labels are in fact physically different situations, even if each
situation individually is empirically inaccessible.  This demonstrates that the permuting of the labels is not a mere descriptive redundancy of the same situation. The only way to evade this is to appeal to a full-blown antirealism by asserting that neither the t-channel nor the u-channel qualify as processes or situations; but in that case, neither direct product ket (in the author's notation, $|\psi_A\rangle_1 \otimes |\psi_B\rangle_2$ nor  $|\psi_B\rangle_1 \otimes |\psi_A\rangle_2$ )  refers to a situation. If neither ket refers to a situation, then neither does their sum (i.e., a symmetrized state). On that view (for example) the singlet state of electrons in a Helium atom does not refer to anything. This is a reductio of such an attempt to deny that the direct product states refer in order retain the conventional claim of redundancy or exchange degeneracy. In effect, the claim becomes vacuous at best, since of course quantities that all refer to nothing are ``degenerate,'' but only in a trivial sense.

It is worth emphasizing that the antisymmetry of the fermion states results from the interchanging of Fermion field operators required to take into account the two interaction channels--i.e. the two different ways that the incoming states can transition to the outgoing states. Based on the Feynman diagrams, these channels are first order transition amplitudes from initial direct product states to the final direct product states (eqs. 3 and 4).\footnote{Of course, this is a perturbation expansion and there are higher-order contributions, but the first-order contribution is the major effect in typical scattering situations. It is also the one often pointed to as an example of the need for symmetrization, e.g. as in this tutorial: https://web.physics.wustl.edu/~wimd/Q540-17-01.pdf. We also focus on the first-order diagrams since they are the physically correct version of the two cases depicted in Bigaj (2022, 225).}
 The interchanging of the field operators needed to take into account each transition amplitude (i.e. the t- and u-channels) introduces a minus sign based on their anticommutation relations; we see this in the sign change of (4) in which the primed (outgoing) momentum indices are swapped. Thus, symmetrization really applies not to a particular state, but to the total transition amplitude from the initial direct product states (typically, specific momentum eigenstates) to the final states, and it is actually a kind of shorthand to apply symmetrization directly to the states themselves.

 To clarify, the arbitrary labels appearing in symmetrized states function as surrogates for variables that can take on specific measurement outcomes such as ``momenta $p_A$ and $p_B$'' that are actually detected. The latter apply to the outgoing electrons in scattering, where each outcome could have come from either incoming state such that those amplitudes add. For the case of bound electrons such as in a Helium atom in a particular eigenstate of total momentum (singlet or triplet), no outgoing state is yet specified since the electrons remain bound, but the potential for measurement outcomes is reflected in the typical wavefunction representation, where the indices are imported as position variables; e.g., $\Psi(x_1, x_2)$.  Wavefunctions are amplitudes, underscoring the above point. Were a measurement actually conducted, yielding outcomes such as (A,B;  spin value), we would then have a factorized state, just as in the free particle case above. At that point, there is no need for the arbitrary labels. Instead they are distinguished by their measurement outcomes. (One might note that this is actually harmonious with Bigaj's criterion for distinguishability; e.g., Bigaj, (2022, 191). All he need do is take into account that the arbitrary labels need not be viewed as permanent tags requiring ongoing formal symmetrization, but instead as surrogates for specific measurement outcomes.)

 The tradition of attributing symmetrization to states as opposed to amplitudes obscures the essential physics of the exchange interaction, in which symmetrization really applies to the total transition amplitude, and any entanglement persists only until measurement yielding an outcome-eigenstate. This point underscores the fact that the conventional treatment of symmetrization as a universal postulate needed to overcome an alleged degeneracy of direct product states is detached from the actual physics of the interactions taking place, which in general concerns transition amplitudes and which (in the real world) goes beyond first-order. If one were to take into account higher order effects, there would be appropriate additional contributions not limited to conventional symmetrization (e.g. Banerjee {\it et al}, 2022).
 
 Another observation is in order regarding the crucial role of the exchange interaction. A typical discussion  (e.g. as in Hutem and Boonchi 2012) starts with a symmetrized Helium eigenstate and evaluates the expectation value of the Coulomb perturbation $ \frac{e^2}{| {\bf r_2} - \bf{r_1}|}  $ with respect to that state. But the interaction is general and also applies to a situation involving free electrons in well-defined incoming states; it is the interaction that couples those states, and that is what is expressed in the above scattering situation and in (2). The latter can be more directly seen from the application to ferromagnetism in Jeschke (2018). 
 
 Neutral fermions, such as neutrons, remain subject to the electromagnetic interaction through their intrinsic magnetic moment as well as their nonuniform internal charge distribution (cf. Nowakowski, Kelkar and Mart, 2006). Thus, even for a 'neutral' particle such as the neutron, the electromagnetic interaction cannot
be switched off.\footnote{While the physical status of neutrinos remains a matter of study, any process that annihilates incoming neutrino states and creates outgoing ones would serve to underlie a symmetrizing correlation. This includes self-correlations of a fermionic field under conditions in which these are non-negligible, and thus
 the term ``interaction'' does not necessarily mean a distinct field mediator (such as the Coulomb field).} 

There have been claims of experimental demonstration of neutron anti-bunching for non-interacting neutrons, e.g. Iannuzzi et al (2006). 
These authors use a beam of thermal neutrons and obtain data indicating anti-bunching. However, they presuppose that the neutron beam is Poissonian, which a priori neglects any possible interactions among the neutrons, however small (or between neutrons and the source/apparatus components). Then, since they observe a small anti-bunching effect, they have nothing else to attribute it to besides formal anti-symmetrization. A more careful analysis would take into account at the very least the conditions of neutron emissions from the source material, the neutron's magnetic moment and also its interaction with the beam splitter, which could involve correlating effects. As a case in point, another experiment by Basar et al (2009) shows non-Poissonian behavior in thermal neutrons. While the source in that experiment is a crystal, the authors in the anti-bunching experiment (Ianuzzi et al) do not discuss their source or process in obtaining the neutron beam. They appear to assume that it produces a Poissonian beam without showing that there can be zero source-related correlations. The basic point is that experiments evidencing anti-bunching in neutral fermions do not support a requirement of a priori formal symmetrization unless it has been shown that neither particle interactions nor self-correlations can possibly be significant enough to result in the anti-bunching observed.  
 
Let us recall again the convention of assigning symmetrized states to all fermions entering experimental situations even if they have not
interacted (or self-correlated, e.g. via the pair correlation $g^{(2)}$) in any way. This convention is often justified by invoking abstract ``symmetry properties'' of fermions. But we have seen in the above scattering example that this initial formal symmetrization is superfluous: it does not enter into the actual calculations of the scattering 
amplitudes, since only one specific field state is destroyed or created at any particular vertex. The electrons do gain a collective antisymmetric
state due to their interaction, which involves specific permutation of the relevant field operators in setting up the two distinct channels.
But this collective state is not based on some abstract ``symmetry property'' that fermions carry with them regardless of the processes in which they participate. Rather,
the commutation relations become relevant through specific physics, such as the two scattering channels above, which together influence the final 
collective state.

The convention of assigning symmetrized states to independent fermions, which is being critiqued herein, seems justified and persists because it serves as a computational stand-in for the actual physical interactions taking place. We see this again in the discussion of an experiment by Neder et al (2007) involving interference effects between electrons in identical spin states (Figure 5).

\begin{figure}[h]
   \centering
    \includegraphics[width=0.4\textwidth]{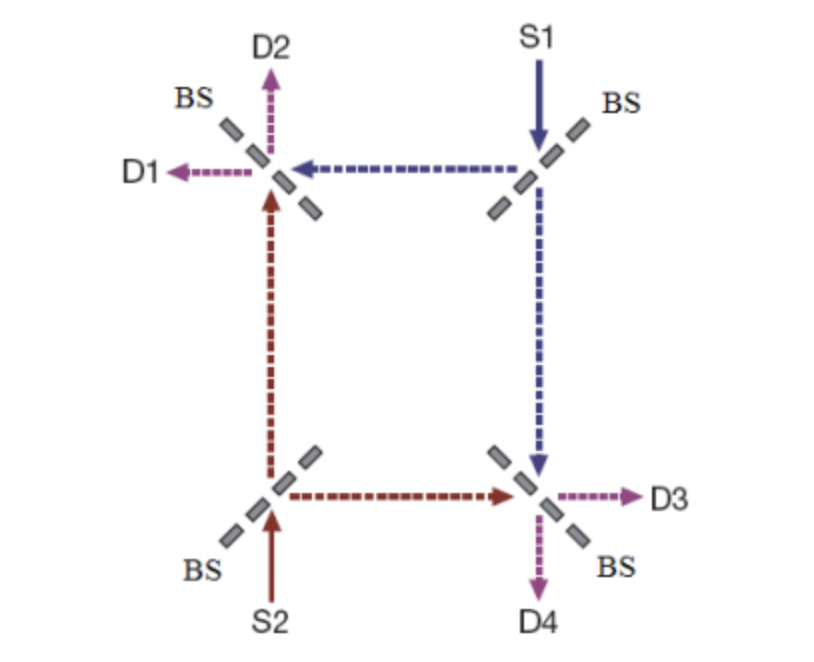}
    \label{fig:your-label}
\end{figure}

\begin{figure}[h]
   \centering
    \includegraphics[width=0.4\textwidth]{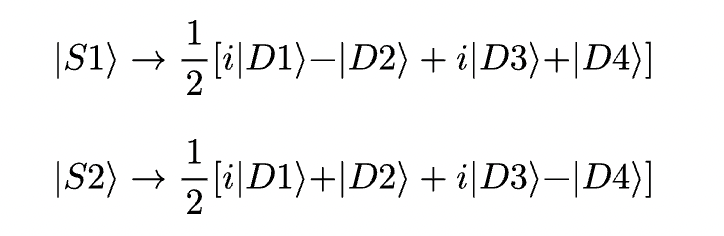}
    \caption{Interference of two electrons and evolution of each source state}
    \label{fig:your-label}
\end{figure}

 The authors send electrons from independent sources into a double-interferometer setup such that the two electrons' wavefunctions can overlap, and find the expected fermionic behavior (e.g. no two electrons ever arriving at the same detector). They attribute this behavior to ``quantum exchange statistics,'' which is the conventional interpretation.
The authors derive the expected results by assigning a collective antisymmetrized state to the two electrons prior to their entering the interferometer and then
evolving this state under an assumed trivial Hamiltonian (similarly to the treatment of Bigaj). In other words, they neglect any interaction/scattering between the electrons. But in fact, the experiment is a scattering situation, where the scattering centers are areas in which the wave functions
overlap. Under these conditions, M\o ller scattering applies. One can analyze the experiment this way, without formally antisymmetrizing the incoming
electron states, and obtain the same results. One finds that physically justified antisymmetrization arises through taking into account the M\o ller scattering channels
and their relative minus sign. In effect, assigning an antisymmetric state to the two electrons prior to their entering the interferometer acts as 
a substitute for the t- and u-channel scatterings in which the electrons actually participate (but which is neglected in applying trivial
propagation to the incoming state). Consider for example the amplitude for detection of both electrons at the same detector, which is zero. One gets this straightforwardly
without symmetrizing ``in advance'' by looking at the applicable scattering amplitude: that scattering process involves identical t and u-channel sub-amplitudes with a mutual minus sign, so that one has destructive interference and the total amplitude is zero. (Recall that the electrons are in identical spin states.)

 In this regard it is worthwhile to recall the comment of Nobel Laureate R. Glauber (1995): ``The things that interfere in quantum mechanics are not particles. They are probability amplitudes for certain events. It is the fact that probability amplitudes add up like complex numbers that is responsible for all quantum mechanical interferences.'' The point is that it is not abstract ``statistical symmetry properties'' that dictate physical phenomena such as the two electrons never being found at the same detector, but specific interactions among probability amplitudes. It is the latter that reflect whatever phase relations are in play due to the commutation relations of the field operators.

 One might object that fermions can exhibit antisymmetric correlations in the absence of a specific interaction involving an additional field such as the electromagnetic field. Indeed, a particular fermionic field may have self-correlations involving antisymmetric effects, such as those discussed above arising from the pair correlation function $g^{(2)}$. However, this is generally a short-range effect and clearly is not significant enough to influence M\o ller scattering; we have seen that the antisymmetric form is fully accounted for by two scattering channels arising in the second-order S-matrix interaction term. The point is that specific anti-symmetric behaviors must be described by specific physical processes involving relevant applications of the field operators, whether through an additional field or through self-correlations. This applies not just to fermions but to bosons, as we briefly address in Section 5.

\section{``Redundancy'' Orthodoxy Neglects Crucial Physics}

  The above counterexample  of M\o ller scattering demonstrates that both the approaches mentioned above--essentialism and the traditional haecceitistic approach--neglect crucial physics that constitutes a meaningful distinction between the permuted states (even if not an empirically accessible one). Thus, arguably there is in fact no redundancy, and accordingly no real degeneracy problem that needs to be avoided or remedied through the postulation of symmetrization as a universal formal rule. Instead, the direct product summands in entangled symmetrized states are indeed individually physically meaningful; specifically, they refer to distinct processes that must be combined in the appropriate phase--however unobservable each such process might be. This points to a need to critically evaluate the implicit assumed equivalence between ontology and `observational effects', as evidenced in this comment from the above extract from Bigaj (2015):  ``but this difference cannot give rise to any observational effects, as haecceities are not empirically accessible.'' Here, the author presupposes, without argument, that any physically relevant difference must be one that ``gives rise to observational effects.''  That is, in the conventional debate on both sides, {\it the physical relevance of any ket (including a direct product ket) is tacitly assumed to require that it be independently associated with specific observational effects}. This assumption is directly challenged by the scattering counterexample presented above: the processes described by the permuted kets are physically distinct, and both are evidently necessary---in the appropriate phases---for ultimately correct empirical correspondence. (It may also be worth pointing out that the evident conventional tacit criterion of observability for any physically relevant feature or process is reminiscent of Mach's rejection of Boltzmann's postulation of unobservable atoms as the basis of thermodynamical phenomena. History records which approach was the more fruitful in that debate; without Boltzmann's proposal, we would not have statistical mechanics, quantum mechanics, atomic physics, or essentially any of modern physics.)
  
  Thus, the current work appeals for a critical awareness of the underlying, largely unexamined empiricist assumption underlying the conventional debate--specifically, that only that which directly corresponds to an observational effect should be considered physically real or at least physically relevant--which persists on both sides of the current debate between what Bigaj refers to as  ``orthodoxy'' versus his own``heresy''. (The ``orthodox'' camp in Bigaj's formulation is exemplified by the work of French and collaborators; e.g., Krause and French (2007).)
 
In any case, in reference to the boldfaced passage quoted above, one cannot term the permuted states ``empirically indistinguishable'' since in fact (for entangled systems) such states are  empirically {\it  inaccessible}, and these two terms are not equivalent (Kastner 2023). Specifically, if one has no empirical access to either situation A or situation B, such that one has associated empirical phenomena for neither A nor B , then when cannot say that these situations are empirically indistinguishable. For the latter implies that A and B instantiate the same empirical phenomena, which is not the case, since they instantiate {\it no} empirical phenomena (for entangled systems). Despite that fact, the conflation of ``empirically inaccessible'' with ``empirically indistinguishable'' seems to have become standard practice in the conventional debate (and arguably this has led it astray).

With this background, let us take a closer look at the notion of ``factorism'' mentioned in Section 1. Bigaj defines ``factorism'' in terms of the assertion that the individual Hilbert spaces appearing in direct products ``represent states and properties of one individual particle'' (Bigaj 2022, 32), but the dependence of this definition on the notion of ``properties'' makes it somewhat ill-defined. For example, one cannot necessarily identify such `properties' with eigenstates of observables; but to view them as ``beables'' is not necessarily justified either. From the context, however, it seems clear that Bigaj and Caulton reject any specific physical meaningfulness of non-symmetrized direct product states. Bigaj seems to allow some vague room for physical meaningfulness in certain comments such as ``Operators which represent properties of individual particles are meaningful but, strangely enough, they are not literally observables.” (Bigaj 2015, 60). As noted in Kastner (2023), this only seems ``strange'' under the empiricist assumption that ontology should be equivalent to observability. This slide is further evidenced in Bigaj's argumentation that the direct product states cannot actually be ``occupied'' by systems, for example in Bigaj (2022,55) where he says: 

\leftskip0.5cm\relax
\rightskip0.5cm\relax
\noindent
 {\small The quantum case of permutation-based redundancy is different from the classical one...not only are the kets $ |\phi\rangle_1  |\psi\rangle_2 $ and $ |\psi\rangle_1  |\phi\rangle_2 $ distinct; they are also orthogonal, which means—according to...the Born rule—that the probability of finding the system in one state given that it occupies the other one should be zero. So it seems that the inclusion of both kets in our representational framework leads to a logical contradiction.}
 
 \leftskip0cm\relax
\rightskip0cm\relax
\hspace{0pt}\\
 \noindent However, a contradiction only arises under Bigaj's presupposition that such states should represent the same physical situation. The scattering example, along with application of the attendant theoretical apparatus, shows that this assumption is undermined by specific physics. It may also be noted that the scattering interaction is a direct counterexample to Caulton's ``second reason'' for denying the meaningfulness of the direct product states. He says:

\begin{quote}

{\small My second criticism of factorism is that it defies an interpretative principle that ought
to be compulsory; namely that the unitary equivalence of two Hilbert spaces and accompanying algebras is a sufficient condition for considering those Hilbert spaces to be
equally good mathematical representations of the same space of physical possibilities. (Caulton 2014, 14)}

\end{quote}

While it is encouraging to see Caulton embracing the notion of physical possibilities as inherently meaningful, it appears
that his mistaken conclusion of redundancy of these possibilities arises from limiting his analysis to the non-interacting Hilbert space. In other words, the algebra of a non-interacting space does not exhaust the relevant physics. To assume that it does amounts to neglecting the exchange interaction, as reminded above. The latter is what results in the need to take into account the above distinct scattering processes and arguably symmetrization in the first place, as implied by (2).
 
 \section{Uncritical empiricism needlessly constrains the space of interpretive solutions}

The uncritical equating of ``empirical'' to ``physical'' and ``meaningful''---i.e., the idea that nothing can be considered physically meaningful unless it is directly associated with an empirical phenomenon---can also be seen in Bigaj's elaboration on the ``argument from exchange degeneracy'' as follows (emphases added):

\begin{quote}

\small{ Suppose that indeed it is possible for a two-element system of same-type particles to occupy states that are neither symmetric
nor antisymmetric, in particular states that are products of two non- identical (orthogonal) vectors. Let us select one such state of the form
$ |\phi\rangle_1  |\psi\rangle_2$ . As we already know, the Hilbert space  $ H  \otimes H $ also contains a
permuted vector$ |\psi\rangle_1  |\phi\rangle_2$ which, { \bf by assumption, represents a state that is empirically indistinguishable from $ |\phi\rangle_1  |\psi\rangle_2$ }. Thus we have a case of what is known as representational redundancy: our mathematical framework contains distinct representations of  {\bf the same physical, or empirical,} situation. (Bigaj 2022, 54)}

\end{quote}

However, again, nobody really is (or at least nobody should be) {\it assuming} that the two states mentioned above are ``empirically indistinguishable'' since (when interference effects are present) they are never individually instantiated.  As noted above, the most that can be said is that they are empirically inaccessible. But it does not follow that these states are not physically meaningful as component states, contrary to the slide in the last quoted sentence that explicitly (and again, uncritically) equates ``physical'' to ``empirical.''  In contradiction to this assumption, we have seen that such component states have distinct well-defined physical referents in connection with the above scattering example.

 Another consideration directly contradicts the idea that the direct product state permutations are ``empirically indistinguishable": If expressions $\psi_{A}(x_1) \psi_{B}(x_2)$  and $\psi_{B}(x_1) \psi_{A}(x_2)$ individually lead to different probabilities, which they certainly can, then they are empirically distinguishable, at least in principle. That is, {\it if} these states were instantiated by real systems, we would find them empirically distinguished. This recalls the point in Kastner (2023) that one standard argument for the need for symmetrization is that the direct product states taken individually are {\it noninvariant} concerning empirical phenomena and thereby empirically distinguishable in principle. (Oddly, Bigaj acknowledges that the individual direct product kets are non-invariant, yet still falls in with the convention that these kets are  ``empirically indistinguishable.'' He says: In particular we can't distinguish two possible final states after detection  $|\psi_A \psi_B\rangle $  and $|\psi_B \psi_A\rangle $, since these kets are not permutation-invariant.'' Bigaj 2022, 241.  But why would we {\it not} be able to distinguish two things that are non-invariant under permutation? Isn't this exactly the opposite of what is meant by non-invariance--i.e., that the two situations are distinct? Apparently, what he means instead is just that the non-invariant states are never actually observed. But again, in that case our inability to distinguish them is only vacuously true: of course one can't distinguish among situations that one cannot observe. I cannot distinguish between a black object and a white object if I cannot see either object. But this does not demonstrate representational redundancy of the terms ``black" and ``white."

Besides the uncritical equating of ``physical'' to ``empirical'', Bigaj's assertion that ``both [direct product] kets are supposed to represent the same empirical situation'' (Bigaj 2022, 55) cries out for critical examination, especially in view of the above point that their absolute squares can be different. Whence such an expectation? As far as this author is aware, there is no such edict or requirement. The locution ``supposed to represent the same empirical situation'' simply expresses the unexamined empiricist denial of a distinction between the kets based on their empirical inaccessibility (an assumption we have refuted herein).

 Thus,  a form of uncritical empiricism has crept into the discussion that serves to preemptively exclude the possibility that the named direct-product states physically refer, and that they refer to distinct physical situations. This exclusion constrains the study of symmetrized states in a way that arguably has led it off track. Basically, the prevailing orthodoxy has us assigning symmetrized states to same-type particles universally without clear physical justification, thereby (besides resulting in vexing interpretational puzzles) introducing a real and arguably unnecessary redundancy in the formalism, to which we now direct our attention. This issue can be introduced by
 way of Bigaj's stated adherence to the prevailing orthodoxy despite his acknowledgment of  its apparent inconsistency with the empirical facts:

\begin{quote}

\small {``More precisely, all fermionic states are formally entangled, since no antisymmetric state can be written in the form of a product of
vectors.... However, this apparent prevalence of entanglement has no support in experimental facts. For instance, we do not
observe non-local correlations connecting all electrons in the universe. Thus there is a need to come up with a new concept of entanglement that
would be better suited to the task of describing systems of same-type particles. (Bigaj 2022, 153)}

\end{quote}

It should again be noted that there are disagreements about whether symmetrization always involves entanglement, where the latter involves
nonlocal correlations. For example,  Ghirardi, Marinetti, and Weber (2002) argue that a product
state in Fock space, although formally written in terms of Hilbert space as an antisymmetric superposition of product states, remains unentangled. However, Gisin's result (Gisin, 1991), which shows that any state not expressible as a product state implies violations of Bell's inequality, would appear to cast doubt on this (although it may be restricted to single degrees of freedom, such as either spatial or spin states rather than full spinor states). Thus, introducing symmetrization without any specific interaction among the particles may either introduce redundancy (just repeating the same product state with indices swapped) or possibly a nontrivial correlation where there may not actually be one.

In any case, Bigaj mentions one attempt to retain formal symmetrization without entanglement as he sees it; namely, the `pseudo-singlet state' in which the particle with spin `up along z' has location L and the particle with spin `down along has location R:

$$|\Psi_{pseudo-singlet}\rangle = \frac{1}{\sqrt 2} (|\uparrow_z\rangle_1 | R_1 \rangle  \otimes |\downarrow_z\rangle_2 |L_2\rangle  -
 |\downarrow_z\rangle_1 |L_1\rangle  \otimes |\uparrow_z\rangle_2 | R_2 \rangle) \eqno(4)$$

In this construction, symmetrization becomes an artificial procedure dutifully entered into
under the orthodox edict that ``all same-type particles must always be in symmetrized states''. Yet, if the particles really are unentangled, a physically sufficient description would be the simple product state $|\uparrow_z \rangle |R\rangle \otimes  |\downarrow_z \rangle |L\rangle $ where it is acknowledged that this result arises from measurement with the corresponding outcomes.  Of course, this is a heterodox approach, and despite their own claim that Fock product states are unentangled, Ghirardi et al retain the orthodox belief that ``the only allowed states for a system of two identical particles must exhibit precise symmetry
properties under the exchange of the two particles.''  But we are making the case herein that such a rule is only physically supported in situations where labels become necessary in order to describe the physics, i.e., in the context of an interaction or specifically consequential field correlation  (such as the pair correlation function where physically relevant).  In the absence of such a situation, there is nothing to `exchange': one electron is in the state $ \vert \uparrow_z, R\rangle $ and the other is in the state $|\downarrow_z, L\rangle $. 
The electrons do not possess ``labels'' and never actually did. The labels were just bookkeeping devices for representation of the intermediate possibilities between preparation and detection. Or, they might be viewed as a formal statement that the field creation operators
in an expression such as  $ a_k^{\dagger} a_l^{\dagger}  |0\rangle $ do not operate in any well-defined order. But that information is not retained in actual calculations such as the ones in scattering amplitudes as discussed above. All that is reflected in the amplitude calculations is that one particle enters
the scattering center in some state  $ a^{\dagger} _{p_a,\alpha} |0\rangle $ and the other in some orthogonal state $ a^{\dagger} _{p_b,\beta} |0\rangle $. Then
their indistinguishability as excitations of the same field enters not in permutations of individual Hilbert space labels, but in permutations of specific field operators
corresponding to the specific states in play at the designated vertices.

Thus, given that a measurement (or ``preparation'') occurred which confers specific outcomes on the particles, it is no longer necessary to use numerical labels for the particle subspaces, since those labels were just surrogates for a possible measurement result, as discussed in Section 2. One cannot respond with ``but that direct product state is not allowed,'' since indeed (despite the usual edict repeated by GMW above) it effectively enters into specific amplitude calculations, as noted above. This fact points to yet another reason to reject the idea that formal symmetrization  of states should be a universal postulate; we don't actually need to use it in the lab to correctly account for experimental phenomena. The numerical labels only enter when one needs to take into account several possible unmeasured transitions (channels) from the initial to final free states, e.g. as required by the interaction (2). 

Of course ``measurement'' --capable of yielding definite outcomes such as a well-defined Dirac wavefunction--is a fraught subject in the orthodox approach (and that is an aspect of the problem in accounting for disentanglement
of same-type particles). Nevertheless, one cannot even assign such determinate properties as in the ``pseudo singlet''  to individual particles in the absence of a measurement yielding an outcome (and that is why they are stipulated in Bigaj's proposal, which is a variant of orthodoxy). (Kastner (2023, 13-14) mentions a well-defined process of measurement that can serve to disentangle same-type particles, along with localization.) 

Another example of the symmetrization postulate being arguably unsupported by real laboratory procedures and results can be found in the context of Bigaj's unsymmetrized fermion singlet state of the form:

$$|\Psi_{unsym-singlet}\rangle = \frac{1}{\sqrt 2} (|\uparrow_z\rangle_1 | L_1 \rangle  \otimes |\downarrow_z\rangle_2 |R_2\rangle  -
 |\downarrow_z\rangle_1 |L_1\rangle  \otimes |\uparrow_z\rangle_2 | R_2 \rangle) \eqno(5)$$
 
Bigaj contrasts this with his ``pseudo-singlet'' (4) in noting that it has the required spherical symmetry and would therefore produce the spin anti-correlations for arbitrary measurement axes, unlike the pseudo-singlet. But he quickly dismisses (5), noting that it is not symmetrized, and proceeds to substitute the usual product of symmetrized space state and anti-symmetrized spin state as the properly symmetrized singlet state.  However, this raises the following issue: 
if Nature really requires all identical particles to be always symmetrized, the state (5) should not be available to real electrons. In other words, acceptance of SP implies acceptance of the proposition that the state (5) can never obtain in Nature. Yet there seems
to  no reason that such a state could not be prepared (at least as an initial state before correlating interactions commenced, i.e. as an input state to a scattering region).

The above considerations raise questions about empirical consistency of the universal postulation of symmetrization. But returning 
to our concern about the arguments invoked for this postulate, involving alleged ``redundancy'': much of the debate has been premised on an untenable assumption of physical meaninglessness
or ``representational redundancy'' concerning mathematical objects in quantum theory that may well have important physical content. Another example of this sort of unwarranted cancellation of theoretical content is Bigaj's assertion that phases of quantum states constitute ``surplus structure'':

\begin{quote}
{\small Representational redundancy is a common occurrence in mathematical physics, thanks in part to the richness and flexibility of mathematical
formalism (which implies in the majority of cases the existence of so- called surplus structures, in Redhead’s terminology, see Redhead 2002). A
well-known case of redundancy present in quantum mechanics is caused by the fact that two vectors that differ by a phase represent the same
physical state. (Bigaj 2022, 54)}

\end{quote}

But in fact, this assertion depends on a very restricted and arguably deficient reading of what counts as a ``physical state.'' The Aharonov-Bohm effect---an empirically verified phenomenon---occurs precisely because of phase changes accumulated by the relevant quantum states, and the influence of forces governed by interaction Hamiltonians are effected by relative phase changes in the components of superpositions. Thus it cannot be claimed based on the fact that, e.g., global phases cancel out in calculating an expectation value, that phases are mere mathematical, descriptive redundancies. The latter is part of an unnecessary and unwarranted deflationary ``what you see exhausts what exists'' tradition towards the theoretical content relevant to same-type particles and quantum systems in general.

Moreover, Bigaj's analysis (Bigaj 2022, Section 3.2) shows that it is only under the arguably mistaken assumption of permutation invariance (based on ``exchange degeneracy'')  that the universal postulation of symmetrization becomes necessary. Ironically, that section shows that Bigaj's essentialist interpretation of the exchange of indices leads to contradictions, which are only remedied  through an {\it ad hoc} symmetrization postulate (SP).  Thus the SP only arises to cure the false problem arising from inappropriately taking the direct product kets as representing the same physical situation. In fact, we clearly see in the correctly rendered scattering situation (Figure 1)  that the two distinct kets do not represent the same physical situation, so there is no actual exchange degeneracy, and thus no need to impose symmetrization as an {\it ad hoc} postulate to cure the contradictions arising from the assumption of exchange degeneracy. Instead, symmetrization is physically required by the need to have both processes included and the fact that neither is more physically important than the other. In this understanding, there is no ``surplus structure''. The individual direct product kets are physically meaningful in describing distinct processes, however empirically inaccessible they might be. 

\section{Boson symmetrization}

It is generally argued that boson state symmetrization is mandatory for predicting certain observed bunching effects such as in the Hanbury-Brown-Twiss (HBT) effect
and the Hong-Ou-Mandel (HOM) effect. However, before discussing the conventional view on these phenomena, we should note that Vatarscu (2023) has raised significant concerns as to the applicability of the standard analysis for the HOM effect, which describes photon propagation through a beam splitter as a unitary transformation. He argues that this approach neglects physically important contributions such as loss of photons and Rayleigh emission. Thus, at the outset, it should be noted that there is no firm consensus on how to analyze or interpret these effects. This means that one must apply care when adducing support for any particular interpretive view by reference to this class of effects involving photons. 

Given that caveat, let us consider the standard account of these effects. First, Nazir and Qureshi (2024) have proposed a formulation that unifies the HBT and HOM
effects as involving different orders of interference, where HOM represents the minimum order of interference, and HBT represents the maximal order.  They model
these experiments in terms of a formal beam splitter with n ports, where HOM (pictured in Figure 6) corresponds to $n=2$ and HBT corresponds to $n \rightarrow \infty$. These authors
assume the standard symmetrized states to obtain their results. We mention this just to clarify that one can see the HBT and HOM as having a common origin in 
terms of interference of indistinguishable photons, under the conventional approach. 

 \begin{figure}[!h]
\centering\includegraphics[width=2in]{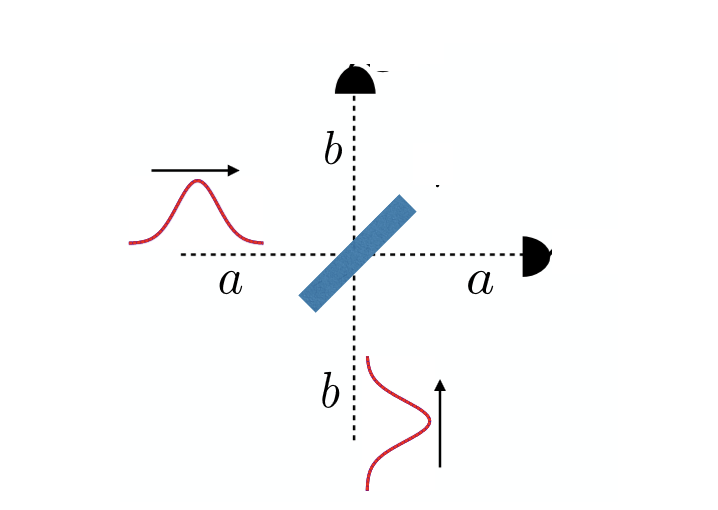}

\label{fig_sim}
\caption{Basic Setup for the Hong-Ou-Mandel effect.}

\label{fig_sim}
\end{figure}

 However, let us recall that the bunching behavior can be predicted based only on evolution of the field operators representing different possible
 physical processes leading to detection of the two photons, depicted in Figure 7:
 
  \begin{figure}[!h]
\centering\includegraphics[width=4in]{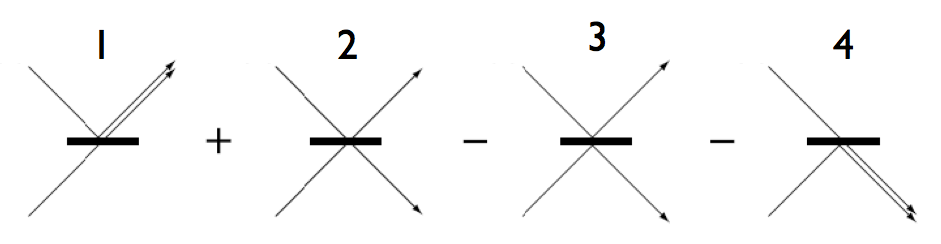}

\label{fig_sim}
\caption{Different possible processes underlying  the Hong-Ou-Mandel effect. The third process is reflection from both surfaces,
in contrast to the second which is transmission through both surfaces.}

\label{fig_sim}
\end{figure}

These processes are reflected (for example) in the analysis in Agata M. Branczyk (2017), Section 3.2.  The photon-bunching prediction is evident in their eq. (11) without evaluating the explicit field state, simply by looking at the evolved field operators: i.e., 

\begin{figure}[!h]
\centering\includegraphics[width=3in]{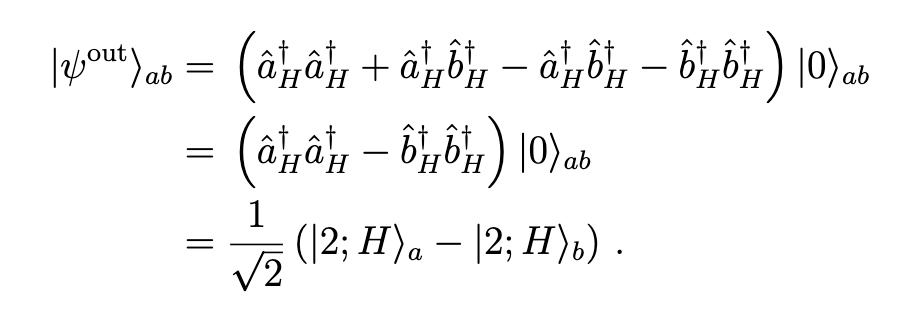}


\end{figure}
\hfill (6) \smallskip

  The photon field operators commute, so the cross terms, which reflect the two different ways in which one photon goes to each of the detectors, end up the same and mutually cancel. Note that if one were to take these two expressions as referring to ``the same physical situation,'' then the question would arise as to why one has two terms with the opposite sign such that cancellation of that ``same physical situation'' takes place. That is, a single physical situation cannot cancel itself. Thus, formal state symmetrization is not the fundamental reason for the bunching effect, which instead arises directly from the commutation relations of the field operators. 
 
  Moreover, for photons in orthogonal polarizations (and thus ``distinguishable''), we don't get bunching regardless of formal state symmetrization, i.e.:
   
    \begin{figure}[!h]
\centering\includegraphics[width=4in]{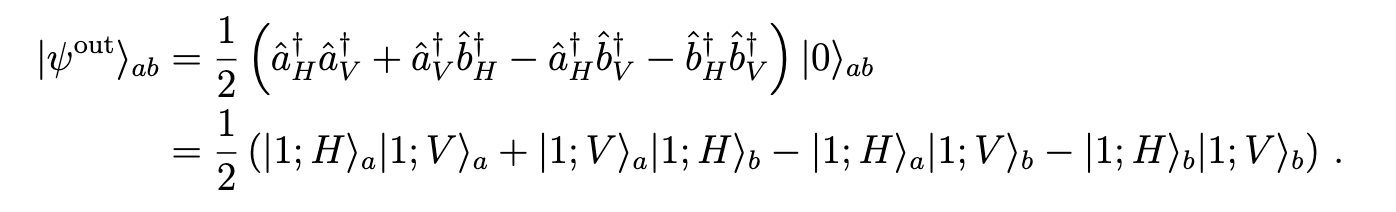}

\end{figure}
\hfill (7) \smallskip

 The prediction for ``distinguishable'' photons (7) does not involve symmetrization at all, since the same result is obtained from
 evolving the input states through the beam splitter in a first-quantized treatment, in which case we get for photons with
 orthogonal polarizations (H and V):
 
 $$|\Psi\rangle = \frac{1}{2} (|a, H\rangle |a, V\rangle  + |b, H\rangle |a, V\rangle - |a, H\rangle |b, V\rangle -|b, H\rangle |b, V\rangle  \eqno(8)$$
 
 This demonstrates that universal symmetrization of boson states is not required for empirical consistency with many observed
 phenomena. In a first-quantized treatment, symmetrization is required only for photons with select identical properties, and then only
 as a computational substitute for the more directly applicable second-quantized treatment.  Thus the formal symmetrization is not about removing a redundancy. All it is 
 doing is acting as a stand-in for neglected applications of the field operators to the relevant physical processes. Formal symmetrization can be used as a `shorthand' if direct calculations with field operators are not used, but again, that still won't yield bunching if the photons are in orthogonal spin states. This shows that formal symmetrization is not necessary or relevant, in general, for empirical consistency. Commutation relations of field operators are indeed involved, but not in the form of an a priori imposition of a symmetrized state. As observed above, the imposition of formal symmetrization does not actually contribute to the bunching prediction in general. It only yields bunching predictions for a subset of photon states, and then only when used as a stand-in for neglected physical processes that are more accurately described by the second quantized treatment.

 The basic point is that it is the appropriate evolution and action of the field operators that does the work of accounting for the collective behavior. When symmetrization is appropriate, it arises because of the need to take into account commutation relations of the field operators in a particular physical context--whether that is a field correlation function, a process involving unitary evolution of several quanta, or an interaction involving an additional force such as electromagnetism. Along with that observation goes the recognition that the permuted objects have distinct physical significance (even if they can interfere!) rather than being different names for the same physical situation, as is evident in the differing sub-amplitude processes of the H-O-M or Hanbury-Brown- Twiss effects.

\section{Conclusion}

It has been argued that symmetrized states of same-type quanta, such as electrons, arise not from ``permutation redundancy'' but from a physical process--one which creates and destroys states of the same quantum field, and which therefore brings into play the commutation relations for the relevant field operators. (This could involve self-correlations of the field.) A specific physical process, M\o ller scattering,  has been shown to be a direct counterexample to the conventional claim that symmetrized states need to be imposed universally based on an alleged redundancy or degeneracy of permuted direct-product states. Rather than ``exchange degeneracy,'' there is a specific physical process demanding an exchange of field operators. Thus, the direct product states of individual particle spaces can indeed be physically meaningful even if they are not individually observable.  The symmetrized sum of states arises because there are more than one equally important ways that the initial and final states can be connected through the specified process occurring between creation and detection of the particles. 

Moreover, the Feynman diagrams for M\o ller scattering, as well as many other scattering processes among same-type particles, assume that incoming and outgoing particles are in well-defined individual states, and predictions based on those calculations are routinely empirically corroborated. This fact, as well as the physically distinct features of the scattering channels, refutes the claim that symmetrization is, or should be, considered universal based on alleged redundancy of the permuted direct product states. It has been noted that it is self-contradictory to assert a redundancy of permuted direct product states when those same states are mutually orthogonal, distinct, and lead to different probabilities. Thus, in fact no {\it a priori} redundancy exists that needs to be remedied by symmetrization. On the contrary, imposing symmetrization on manifestly non-entangled quanta arguably creates redundancy (or possibly even inconsistency if the symmetrized state implies entanglement that has not actually been prepared). This is the case for the so-called ``pseudo-singlet'' state in which there is a fact of the matter that one particle is in state ``Leftward momentum, spin up'' and the other is in the state ``Rightward momentum, spin down''. Artificially constructing a symmetrized state out of this unentangled state simply adds (at best) an unphysical redundancy.

The overall conclusion is that it is the specific physical applicability of commutation relations for the relevant field operators that accounts
for the observed collective phenomena characterizing same-type particles, including those conventionally attributed to formal state symmetrization. We have seen in the example of M\o ller scattering that it is a specific interaction that yields the antisymmetrization between electrons; specifically, two distinct sub-amplitudes corresponding to label permutation. Thus, the label permutation is physically meaningful and does not reflect a redundancy that needs to be remedied through formal symmetrization. 
Rather, formal symmetrization of states acts as a computational stand-in for neglected physical interactions, as shown for the case of M\o lller scattering as it pertains to both free electrons and a specific interferometer experiment. Similarly, in the case of bosons, it is the commutation relations that do the real work, not formal symmetrization of states, since the analysis proceeds without necessary references to any explicit representation of the relevant boson states. What yields the bunching or anti-bunching behaviors are 
applications of the commutation relations to specific physical processes, not an {\it a priori} symmetry of the particles themselves independently of any physical process. Thus, arguably it is a misconception to attribute the observed collective behaviors to some abstract mathematical symmetry property or to ``statistics'' elevated to some universal property of the particle in question. If they do not undergo specific physical processes, there is no physical basis for predicting the behavior in question. That means there is no physical basis for attributing symmetrized states to such systems as a necessary {\it a priori} description. It also means that when symmetrization is appropriate, the component direct product states have meaningfully distinct physical referents, whether or not those are empirically observable.

At a more general level, an effort has been made to disambiguate an equivocation in the literature concerning ``empirical indistinguishability'' of states, where that has been inappropriately attributed to direct product states that are not empirically accessible. It has also been pointed out that much of the existing literature adopts an unexamined empiricist stance in effectively equating ontology to observability. The discussed counterexample shows that situations that may not be individually observable, or lead independently to observable effects, are crucially physically relevant.

 \bigskip

Acknowledgments. 

The author gratefully acknowledges valuable discussions with Federico Holik.

\section{References}

Banerjee P., Engel T., Schalch N.,  Signer A., Ulrich, Y. (2022) ``M\o ller scattering at NNLO,'' Phys. Rev D 105, L031904.

Basar, K.  et al. (2009). ``Correlation effects among atomic thermal displacements in oscillatory diffuse neutron scattering of ZnSe,'' 

Basar, Khairul, Sainer Siagian,  Xianglian, Takashi Sakuma, Haruyuki Takahashi, Naoki Igawa (2009).
``Correlation effects among atomic thermal displacements in oscillatory diffuse neutron scattering of ZnSe,''
Nuclear Instruments and Methods in Physics Research Section A: Accelerators, Spectrometers, Detectors and Associated Equipment, Vol. 600, Issue 1,
pp. 237-239, ISSN 0168-9002,
https://doi.org/10.1016/j.nima.2008.11.037.

Beisert, N. (2026). {\it Quantum Field Theory I.} https://edu.itp.phys.ethz.ch/hs12/qft1/Chapter09.pdf, accessed 5/20/26.

Bigaj, T. (2015). ``Exchanging Quantum Particles,'' {\it Philosophia Scientiae 2015}:(19-1),185-198).

Bigaj, T. (2022). {\it Identify and Indiscernibility in Quantum Mechanics}. Cham: Palgrave-MacMillan. 

Bigaj, T. and Ladyman, J. (2010). ``The Principle of the Identity of Indiscernibles and Quantum
Mechanics,'' {\it Philosophy of Science 77}, pp. 117 - 136.
DOI: https://doi.org/10.1086/650211

Branczyk, Agata (2017). ``Hong-Ou-Mandel Interference,'' https://arxiv.org/pdf/1711.00080

Caulton, A. 2014. Qualitative Individuation in Permutation-Invariant Quantum
Mechanics. arXive: 1409.0247v1 [quant-ph].

Caulton, A. (2018): ``Qualitative Individuation in Permutation Invariant Quantum Mechanics,''
arXiv:1409.0247v1.

Cohen-Tannoudji, C., Diu, B., \& Laloe, F. (1977). Quantum Mechanics. London, Paris: Wiley,
Hermann.

Dieks, D. and Lubberdink, A. (2022). ``Identical Quantum Particles as Distinguishable Objects,'' {\it Journal for General Philosophy of Science / Zeitschrift f\:{u}r Allgemeine Wissenschaftstheorie 53}, (3):259-274.

Eisert, J. 2024. {\t Advanced Quantum Mechanics.} https://www.physik.fu-berlin.de/en/einrichtungen/ag/ag-eisert/teaching/ws18-19/AdvancedQuantumMechanicsChapter5.pdf, accessed 5.30.26.

Fano, O. (1961). ``Quantum Theory of Interference Effects in the Mixing of Light From Phase-Independent Sources,'' Am. J. Phys. 29, 539–545.

Ghirardi, G., Marinatto, L. and Weber, T.  (2002). Entanglement and Properties of Composite Quantum Systems: A Conceptual and Mathematical Analysis. Journal of Statistical Physics 108, 49–122. https://doi.org/10.1023/A:1015439502289

Gisin, N. (1991) Bell’s Inequality Holds for All Non-Product States. Physics Letters A, 154, 201-202.
https://doi.org/10.1016/0375-9601(91)90805-I

Glauber, J. American Journal of Physics, 63(1), 12 (1995).

Hutem, A. and S. Boonchui (2012). ``Evaluation of Coulomb and exchange integrals
for higher excited states of helium atom by using
spherical harmonics series,'' {J Math Chem 50}:2086-2102.
DOI 10.1007/s10910-012-9997-6

Jaouni, T., Gu, X., Krenn, M., D'Errico, A. and Karimi, E. (2025), Tutorial: Hong–Ou–Mandel interference with structured photons. Nanophotonics, 14: 4163-4175. https://doi.org/10.1515/nanoph-2025-0034

Kastner, R. E. (2022). {\it The Transactional Interpretation of Quantum Mechanics: A Relativistic Treatmennt.} Cambridge: Cambridge University Press.

Kastner, R. E. (2023). ``Quantum Haecceity," Philos Trans A Math Phys Eng Sci (2023) 381 (2255): 20220106.

Krause, D. and French, S. (2007). ``Quantum sortal predicates.'' Synthese 154 (3):417 - 430.

Jesche, H. (2018). https://www.physics.okayama-u.ac.jp/jeschke\textunderscore homepage/AP2018/chapter5.pdf. Accessed 12/20/25.

Nazir., K. and Qureshi, T. (2024). ``Generalized Two-Particle Interference,'' https://arxiv.org/html/2404.18468v2

Neder, I. et al. Nature 448, 333‐337 (2007)

Nowakowski M, Kelkar NG, and Mart T (2006). ``Neutron structure effects in the deuteron and one neutron halos,'' {\it Phys Rev C} 74(2); DOI:10.1103/PHYSREVC.74.024323

Schlichtholz, K.  and Markiewicz, M. (2023). Generalization of Gisin’s Theorem to Quantum Fields: https://arxiv.org/pdf/2308.14913

Vatarescu, A. (2023). ``The Quantum Regime Operation of Beam Splitters and Interference Filters,'' Quantum Beam Sci. 2023, 7(2), 11; https://doi.org/10.3390/qubs7020011

Also: zeroth order contribution to already-existing fermions: they don't scatter at all, even if symmetrized:  https://edu.itp.phys.ethz.ch/hs12/qft1/Chapter09.pdf

 \begin{figure}[!h]
\centering\includegraphics[width=2in]{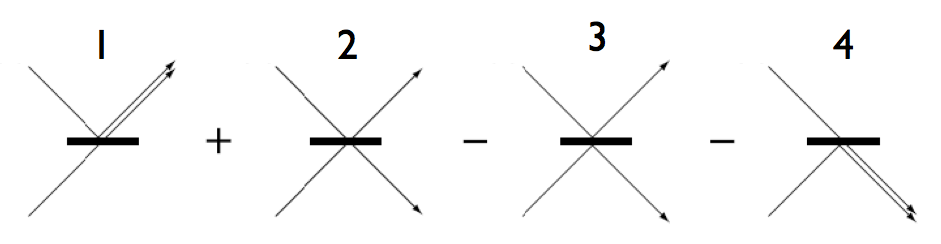}

\label{fig_sim}
\caption{Amplitudes for each of the processes contributing to the Hong-Ou-Mandel effect.}

\label{fig_sim}
\end{figure}

\end{document}